# Patterning and tuning of electrical and optical properties of graphene by laser induced two-photon oxidation

Jukka Aumanen[1], Andreas Johansson[2*], Juha Koivistoinen[1], Pasi Myllyperkiö[1], Mika Pettersson[1*]

[1]Nanoscience Center, Department of Chemistry, P.O. Box 35, FI-40014, University of Jyväskylä, Finland.
[2]Nanoscience Center, Department of Physics, P.O. Box 35, FI-40014, University of Jyväskylä, Finland

*Correspondence to: mika.j.pettersson@jyu.fi or andreas.johansson@jyu.fi

**Abstract**

Graphene, being an ultrathin, durable, flexible, transparent material with superior conductivity and unusual optical properties, promises many novel applications in electronics, photonics and optoelectronics. For applications in electronics, patterning and modification of electrical properties is very desirable since pristine graphene has no band gap. Here we demonstrate a simple all-optical patterning method for graphene, based on laser induced two-photon oxidation. By tuning the intensity of irradiation and the number of pulses the level of oxidation can be controlled to high precision and, therefore, a band gap can be introduced and electrical and optical properties can be continuously tuned. Arbitrary complex patterning can be performed for air-suspended monolayer graphene or for graphene on substrates. The method works at room temperature in ambient air and no additional processing step is needed. The presented concept allows development of all-graphene electronic and optoelectronic devices and complex circuits with an all-optical method.

**One Sentence Summary:** An all-optical method is developed for patterning and tuning of electrical and optical properties of graphene.

Graphene has high potential for becoming the next generation material for electronics, photonics and optoelectronics (*1,2*). Electrical properties of graphene can be modified by tuning its shape or dimension. Narrow ribbons of graphene lead to opening of a band gap due to quantum confinement effect. Patterning of ribbons has been achieved by nanolithographic methods and by self-organized growth (*3,4*). However, creation of localized states due to disorder in nanoribbon edges is a problem (*5*). More precise control of band gap has recently been achieved by bending graphene on a patterned SiC substrate leading to localized strain, but it is not clear how to make complex patterns from such structures (*6*). Additionally, this method is limited to SiC substrates. Graphene oxide (GO) has a band gap, which can be tuned by controlling the degree of oxidation (*7,8*). Patterning of GO by reduction using a heated AFM tip has been shown to increase conductivity by four orders of magnitude (*9*). Laser heating has been used for modification of electrical properties of GO (*10-13*). However, thermal reduction of GO does not fully recover the excellent electrical properties of graphene. Laser-based ablation patterning of graphene has also been investigated (*14-18*). Fabrication of sub-diffraction limited features, such as ribbons (*14*) has been demonstrated but formation of disordered edges makes it difficult to modify and control electrical properties. Here, we report patterning and controlled tuning of electrical and optical properties of graphene by femtosecond laser induced non-linear oxidation. The method relies on oxidation without ablation or cutting, thus the carbon network is preserved throughout the process. By tuning the level of oxidation, electrical properties of oxidized regions can be continuously tuned while the excellent electrical properties of unoxidized regions of graphene are preserved.

We imaged patterns by four-wave mixing (FWM), which is a nonlinear optical method giving very strong response in nanomaterials (*19*). Previously, we have used FWM to study individual single walled carbon nanotubes (*20*). FWM of graphene was previously measured and an unusually large third order susceptibility, on the order of $|\chi^{(3)}| = 10^{-7}$ esu was obtained (*21*). We used two input beams with wavelengths of 540 and 590 nm for FWM imaging, during which the sample chamber was purged with nitrogen.

Patterning was performed in ambient air using peak intensity of $10^{11} - 10^{12}$ W/cm$^2$. At higher intensities (>$10^{12}$ W/cm$^2$), ablation of graphene took place. Laser patterning and FWM imaging of monolayer graphene suspended over 7 x 7 μm$^2$ holes on copper TEM grid (Graphene Platform Corp.) is presented in Fig. 1. The FWM signal is very strong and there is high contrast to the copper grid. Raman measurements for the sample showed characteristic features of monolayer graphene including higher intensity for the 2D band than for the G band (figure S1) (*22*). In Fig. 1A, a FWM image of graphene is shown before patterning and in Figs. 1B – 1E images are shown after successive irradiation steps aimed at drawing a rectangular pattern. FWM signal from the pattern decreases after each irradiation step, showing that optical properties of graphene change. The patterned area forms a closed loop proving that patterning does not result in cutting, or ablation. This is a very important distinction to the previous laser patterning works of graphene which relied on cutting and removal of material (*14-18*). The sample was stored for three days at ambient conditions between images 1E and 1F, which shows that patterns are stable.

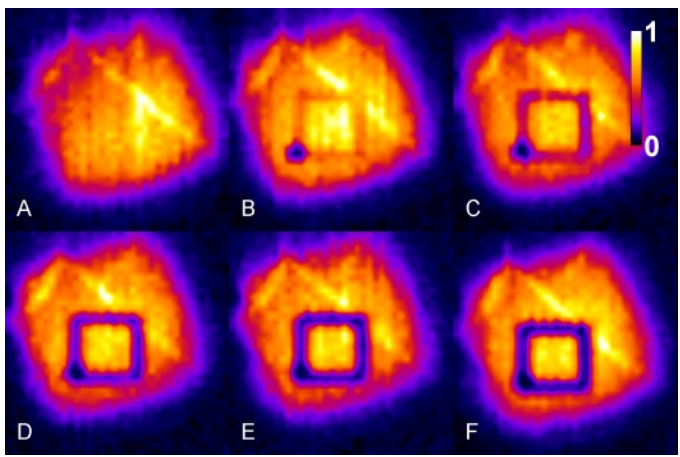

**Fig. 1.** FWM images of patterned suspended monolayer graphene. The image consists of the FWM signal strength at each pixel. (**A**) before patterning. (**B – E**) Patterning in various stages of oxidation. (**F**) Image after storage for 3 days. The spot on the left lower corner is made in purpose in the early stage of patterning. The size of the patterned square is approximately 2 x 2 µm². Graphene is suspended on the backside of the 4 µm thick copper grid which limits focusing of the beam to graphene surface near the edges of the square, thus the size of the FWM image of the graphene square is ~5 x 5 µm² although the actual size of the square is 7 x 7 µm².

Patterning did not occur under nitrogen purge, suggesting that the process involves oxidation of graphene. Fig. 2 presents a FWM image of a sample where two square patches with a size 2 x 2 µm² were patterned, using different irradiation times. Raman spectra indicate that the G-band shifts up and the D-band intensity increases upon irradiation (figure S2), which are signatures of oxidized graphene (*23*). In the lower panel of Fig. 2, the G-band shift along a line marked in the FWM image is shown. The maximal shift is equal to ~13 cm⁻¹, which indicates substantial oxidation as the shift is similar to the observed G-band up-shift between pristine graphite and graphite oxide (*24*). Another indication of strongly oxidized sample is that the intensity ratio of the D- and G-bands (I(D)/I(G)) increased from nearly zero in the non-irradiated area up to a value of 1.6 in the irradiated area (Fig. 2). The I(D)/I(G) ratio is known to gain values between 0.8 – 2.8 in GO and reduced graphene oxide (rGO), indicating possible differing functional group and/or oxygen contents (*25,26*). Observation of characteristic Raman spectrum of oxidized graphene over an extended area, larger than the laser spot size, is another clear proof that no ablation or removal of material is taking place. In order to obtain further evidence of oxidation, three areas of suspended monolayer graphene were irradiated and three control areas were left un-irradiated. Elemental analysis was performed for the areas using energy dispersive x-ray (EDX) analysis (figure S3). O/C atomic ratio is on average 0.09 in the unprocessed areas and 0.22 in the processed areas, confirming oxidation by laser irradiation.

The mechanism of photo-oxidation was studied by measuring kinetics of the decay of the FWM signal at various laser powers for monolayer air-suspended graphene (figure S4). Decay constant as a function of laser power squared is fitted with a linear function ($\chi^2 = 0.98$) indicating that the process involves two-photon absorption ($k \propto P^2$) (figure S5). The effective two-photon cross-section, assuming 100 % quantum yield, for oxidation was obtained from the fit as $4 \times 10^{-54}$ cm$^4$ s (Supplementary Information). This parameter is useful for designing patterning process with various lasers and for estimating the rate of patterning.

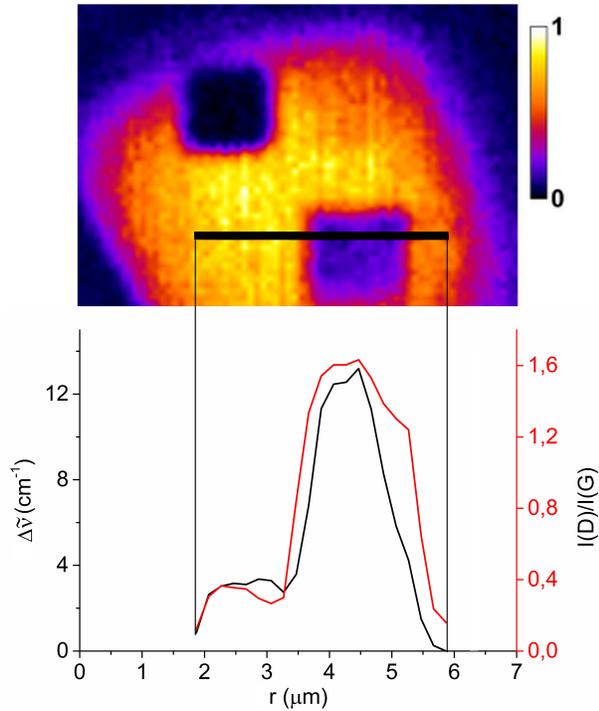

**Fig. 2.** Upper panel: FWM image of a monolayer graphene sample where two rectangular patches were patterned using different irradiation times. Lower panel: Raman spectroscopic data along a line shown in the upper panel. Black line (left axes) indicates the shift of the G-band from its value in non-irradiated graphene and red line (right axis) indicates the value for the intensity ratio between the D-band and the G-band.

It is interesting to compare our results on photo-oxidation of graphene with the previous studies of photoreduction of GO (*10-13*). How is it possible to drive the process into opposite directions by seemingly same method? Photoreduction of GO is based on laser heating, which leads to detachment of oxygen containing groups similarly to reduction by hot AFM tip (*9*). We use only very low average power, on the order of 10 µW, which cannot heat the sample significantly.

Thus, our method is not driven by laser heating but by a process involving excited state chemistry.

We also patterned monolayer graphene on doped Si substrate with a 300 nm thick dielectric layer of silicon dioxide ($SiO_2$) (Graphenea). Fig. 3 shows that features well below a micrometer can be drawn, arbitrary shapes can be made including curved features, the level of oxidation can be tuned and the oxidation level is stable over an extended area. The patterns are also visible in scanning electron microscopy (SEM) (Fig. 3). The line width was determined from the SEM image of letter C (Fig. 3, upper panel) yielding ~400 nm. For comparison, our microscope objective with a numerical aperture of 0.8 yields an estimated focal spot size of 430 nm at 540 nm wavelength. Two-photon excitation makes it effectively smaller by a factor of $2^{1/2}$ yielding ~300 nm, which is in reasonable agreement with the determined line width.

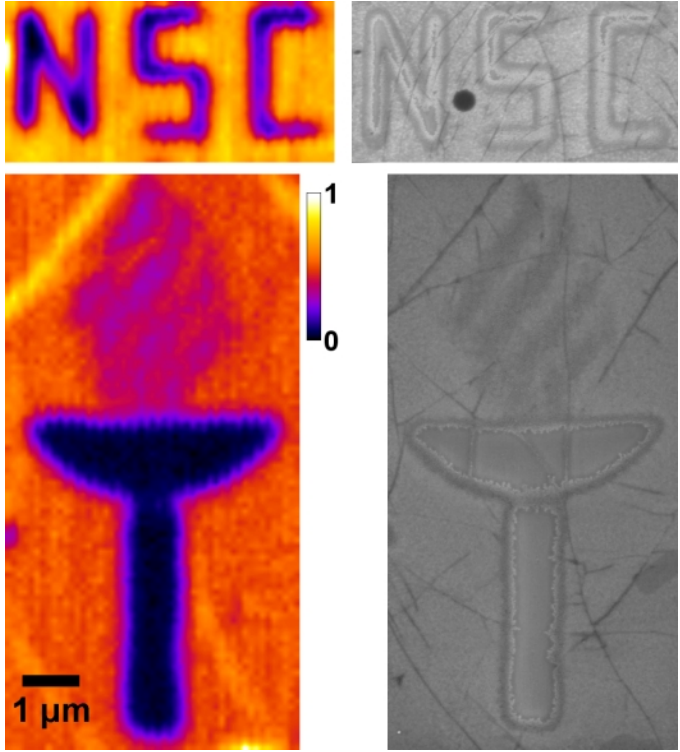

**Fig. 3.** Patterns drawn on monolayer graphene on Si/SiO$_2$ substrate. FWM images are shown on the left and SEM images of the same patterns are shown on the right. The scale bar on the lower left corner applies to all panels.

Electrical properties of the patterned areas were studied. Metal electrodes were fabricated on the SiO$_2$ substrate and the graphene was etched into suitable shapes between the electrodes (upper inset of Fig. 4A). In this structure, two neighbouring electrodes were selected and the I-V characteristics of that section were measured. The I-V was linear with a two-terminal conductance of 15 µS (Fig. 4A). A line between the two electrodes was then gradually oxidized by repeated irradiations (upper inset of Fig. 4A). The I-V curve was measured after each irradiation step and the results are presented in Fig. 4A. The conductance of the graphene sample initially increased with 8.5 percent, after which it decreased gradually with more than 5 orders of magnitude. As the oxidation progressed, the I-V developed a nonlinear shape and at the end turned non-conducting at low voltage bias. The data show the capability to gradually modify the electric properties of graphene through a conductor to insulator transition by changing its oxidation state.

Further evidence of tailoring of the electrical properties by laser patterning was obtained from using the doped Si chip as a back-gate to study the influence of a perpendicular electric field. For moderate laser exposures, when the I-V characteristics still remain linear, no observable change in conductance appears in the back-gate voltage range of -10 V to 10 V. At higher exposure doses, when the I-V has taken on a clearly non-linear shape, also back-gate dependence appears (Fig. 4B). The graphene device acts as a p-type field-effect-transistor, with a current amplification of more than one order of magnitude at negative back-gate voltages. This demonstrates that the oxidation leads to opening of a band gap in the exposed region, which governs the electronic response (*27*). A rough estimate of the band gap can be made from the current onset in the corresponding I-V curve (black trace in the inset of Fig. 4B). By smoothing the noisy I-V curve (red trace) and then differentiating it, the onset appears as a peak in the resulting dI/dV curve (blue trace). The estimate gives a band gap in the range 310-580 meV, due to the I-V not being fully symmetric. This initial result is very promising, considering that in the present case contact resistance is relatively high and the graphene includes grain boundaries, which can be improved in future investigations.

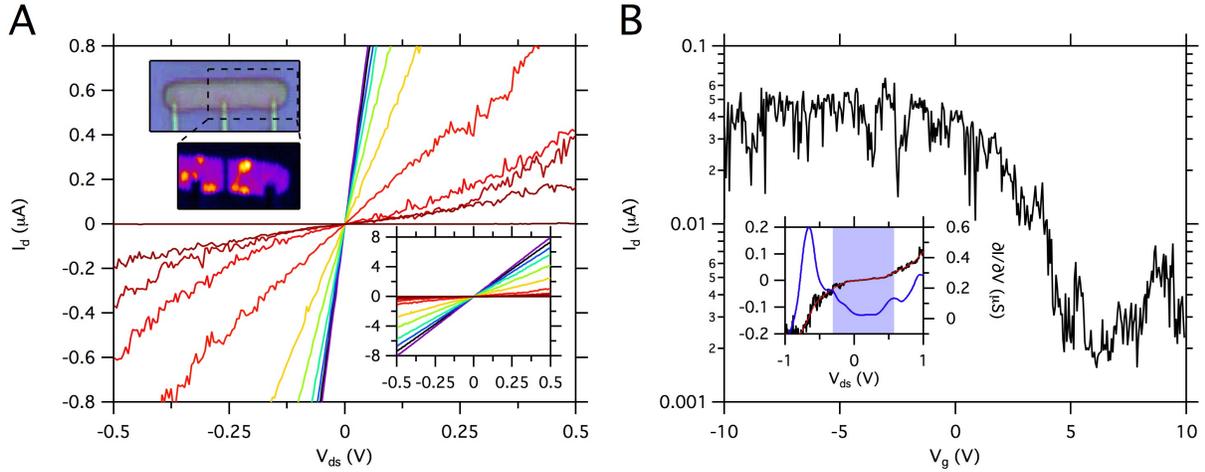

**Fig. 4.** Electronic measurements at different oxidation states. (**A**) Drain current versus applied drain-source voltage, taken before two-photon oxidation (black) and in between oxidation steps (rainbow colors, purple through dark red). The data range captures the change from linear to non-linear I-V characteristics. Upper left inset: Upper pane shows optical image of sample geometry. Distance between electrodes is 6 μm. Lower pane shows a FWM image after the oxidation line is made. Lower right inset: I-V characteristics, displaying the upper range of conductance variation with linear response. (**B**) Current versus gate voltage at a fixed $V_{ds}$ of 100 mV, displaying p-type transfer characteristics. Inset: Corresponding non-linear I-V characteristics (black), the I-V after smoothing (red), and differentiation of the latter trace (blue). The light blue shaded region from -310 mV to 580 mV marks the highly resistive plateau in the I-V trace.

**Acknowledgements.** This work was funded by the Academy of Finland (decision. no. 252468). Janne Ihalainen is thanked for discussions. Hannu Salo is thanked for assistance with EDX measurements.

**Author contributions.** M. P. conceived the idea, planned the experiments, coordinated and supervised the work and wrote the main part of the manuscript. J. A. built the laser setup, performed the laser experiments and contributed to planning of experiments and writing of the manuscript, A. J. designed and built the electrical measurement setup, conducted electrical measurements, fabricated samples, contributed to planning of experiments and writing of the manuscript, J. K. performed Raman measurements and contributed to planning of experiments and writing of the manuscript, P. M. designed the laser setup, developed programs for data acquisition and sample movement for laser patterning and contributed to planning of experiments and writing of the manuscript.

**Materials and Methods**

FWM imaging

The laser pulses for FWM imaging experiments were produced by two non-collinear optical parametric amplifiers (NOPAs, Orpheus-N, Light Conversion) pumped by an amplified femtosecond laser (Pharos-10, 600 kHz, Light Conversion). The two output pulses can be tuned independently from 510 nm to 890 nm and the typical pulse duration in the visible range is ~30 fs. We used 540 nm and 590 nm wavelengths in which case their energy difference matches the G-band frequency of the graphene. The group velocity dispersion induced by transmitting optics was compensated with an additional prism compensator (fused silica prism pair) yielding a pulse duration of ~40 fs at the sample. Relative time delays of the two laser pulses were adjusted with computer controlled optical delay lines (Thorlabs). For imaging, the pulses were overlapped in time. The two beams were attenuated independently with variable neutral density filters before aligning them to collinear geometry by using beam splitters. An additional variable neutral density attenuator was installed in the path of the combined beam in order to adjust the laser power without changing the relative intensity of the two pulses. A tube lens and a camera were installed behind a beamsplitter to view the laser spots on the sample.

Laser beams were focused to the sample by a microscope objective (Nikon LU Plan ELWD 100x/0.80). The sample was installed to a closed chamber that was purged with nitrogen or argon gas during imaging to prevent oxidation of the graphene. The sample chamber was attached on a three-axis piezo-stage (Thorlabs NanoMax 300) to control the position of the sample. FWM signal was collected to backscattering direction and separated from the input beams with dichroic long-pass filter (Semrock) and further purified from the residuals of the input beams by using a bandpass filter (Semrock). Spectrally filtered signal was focused to a photon counting module (single photon avalanche photodiode, SPCM-AQRH-14, Excelitas Technologies). Imaging was performed by a point scan method with typical detection times of 0.1-0.2 s/point. The intensity of the laser radiation at the sample was typically from $5 \times 10^{10}$ to $2 \times 10^{11}$ W/cm$^2$.

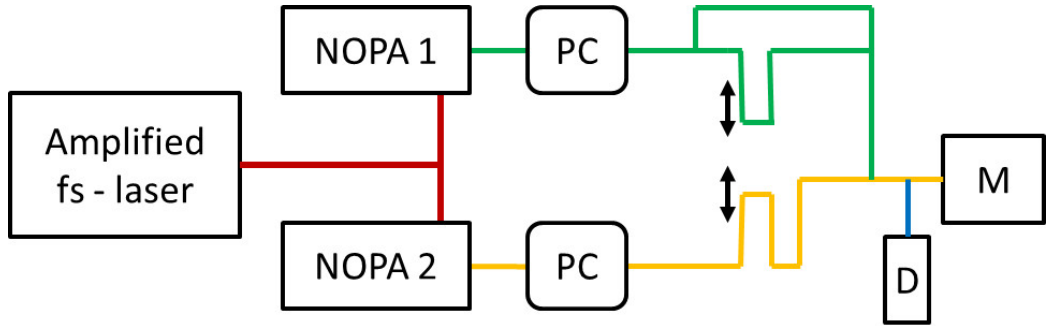

Schematic of the laser setup used for patterning and FWM imaging. NOPA = non-collinear optical amplifier, PC = prism compressor, M = microscope, D = detector.

Photo-oxidation

Local photo-oxidation of the graphene was carried out with the same laser setup as FWM imaging. During oxidation inert gas purge was switched off and the sample chamber contained ambient atmosphere. Patterns were oxidized by moving the sample in 100 nm steps and the position and the irradiation time of each oxidation point was computer controlled. Typical intensities for oxidation were from $1 \times 10^{11}$ to $1 \times 10^{12}$ W/cm$^2$.

Raman spectroscopy

Raman measurements were carried out with a home-built Raman setup (*28*) in a backscattering geometry using 532 nm excitation wavelength produced with CW single frequency laser (Alphalas, Monolas-532-100-SM). The beam was focused to a sample, and subsequently collected, with a 100x microscope objective (Nikon L Plan SLWD 100x with 0.70

N.A.). The scattered light was dispersed in a 0.5 m imaging spectrograph (Acton, SpectraPro 2500i) using 600 g/mm grating (resolution: ~3 - 4 cm$^{-1}$). The signal was detected with EMCCD camera (Andor Newton EM DU971N-BV) using 100 μm slit width. A beam splitter was placed between the objective and the spectrometer in order to observe the exact measurement point visually. The Rayleigh scattering was attenuated with an edge filter (Semrock). The approximate sample positioning was done manually with XYZ-stage (Newport, ULTRAling 462-XYZ-M) and fine – tuned with XYZ-piezoscanner (Attocube, ANPxyz101) with smallest step of 50 nm in each direction. Laser power of ~5 mW was utilized and two 30 s measurements were averaged for each accumulation.

SEM-EDX Measurements

Scanning electron microscope (SEM) measurements and energy-dispersive x-ray (EDX) analysis for carbon (K-electrons) and oxygen (K-electrons), were performed using FEI QUANTA microanalysis system with Zeiss EVO 50 scanning electron microscope and Bruker AXS XFlash Detector 3001. The applied acceleration voltage was 7.53 kV.

Samples for electronic transport measurements

P-doped Si wafers with 300 nm SiO$_2$ and a monolayer of graphene, where the graphene was grown by CVD on Cu and transferred to the wafer, were bought from Graphenea SA, 20018 Donostia-San Sebastiań, Spain. Electrodes of Ti (2nm) and Pd (25 nm) were defined on top of the graphene, using electron beam lithography and physical vapor deposition. Then suitable graphene structures were defined between the electrodes, using electron beam lithography and reactive ion etching (oxygen plasma).

Electronic measurements.

All electronic measurements were conducted at room temperature in ambient conditions. The measurements were computer-controlled, using LabView. The drain-source and gate-source dc voltage bias was supplied through a shielded rack-mountable connector block (National Instruments BNC-2090). The current response through the graphene device was monitored, using a current preamplifier (Stanford Research Systems, SR570). A drain-source bias of 100 mV was used during the transfer characteristics measurements.

**Supplementary Text**

Two-photon cross-section for oxidation

Two-photon cross-section for the photo-oxidation reaction is calculated by using the equations presented by Koester et al. (*29*). We assume that the process consists of two-photon excitation which leads to oxidation with a quantum yield of 1. The cross-section is derived by fitting FWM signal decay constant as a function of laser power squared (Fig. S5). By using the following equation

$$k_2 = \frac{g_p}{\tau f}\left(\frac{\overline{P}\pi(NA)^2}{hc\lambda}\right)^2 \sigma_2$$

where,

$g_p$ = 0.664 (Temporal coherence, We assume gaussian temporal profile)

$h$ = 6.626 x $10^{-34}$ Js (Planck's constant)

$c$ = 2.998 x $10^{10}$ cm/s (Speed of light)

$\lambda$ = 540 x $10^{-7}$ cm (Wavelength)

$f$ = 600 kHz (Repetition rate)

$NA$ = 0.8 (Numerical aperture of the focusing objective)

$\tau$ = 40 fs (Pulse duration at the sample)

$P$ = Measured average laser power at the sample

$k$ = Rate constant at given power from the single exponential fit of the decay of the FWM signal (Fig. S4)

we derive that slope of the linear fit 3.93 x $10^8 = \frac{g_p}{\tau f}\left(\frac{\pi(NA)^2}{hc\lambda}\right)^2 \sigma_2$ and inserting the above listed numerical values we obtain numerical estimation for two-photon cross-section

**$\sigma_2$ = 4 x $10^{-54}$ cm$^4$ s photon$^{-1}$**

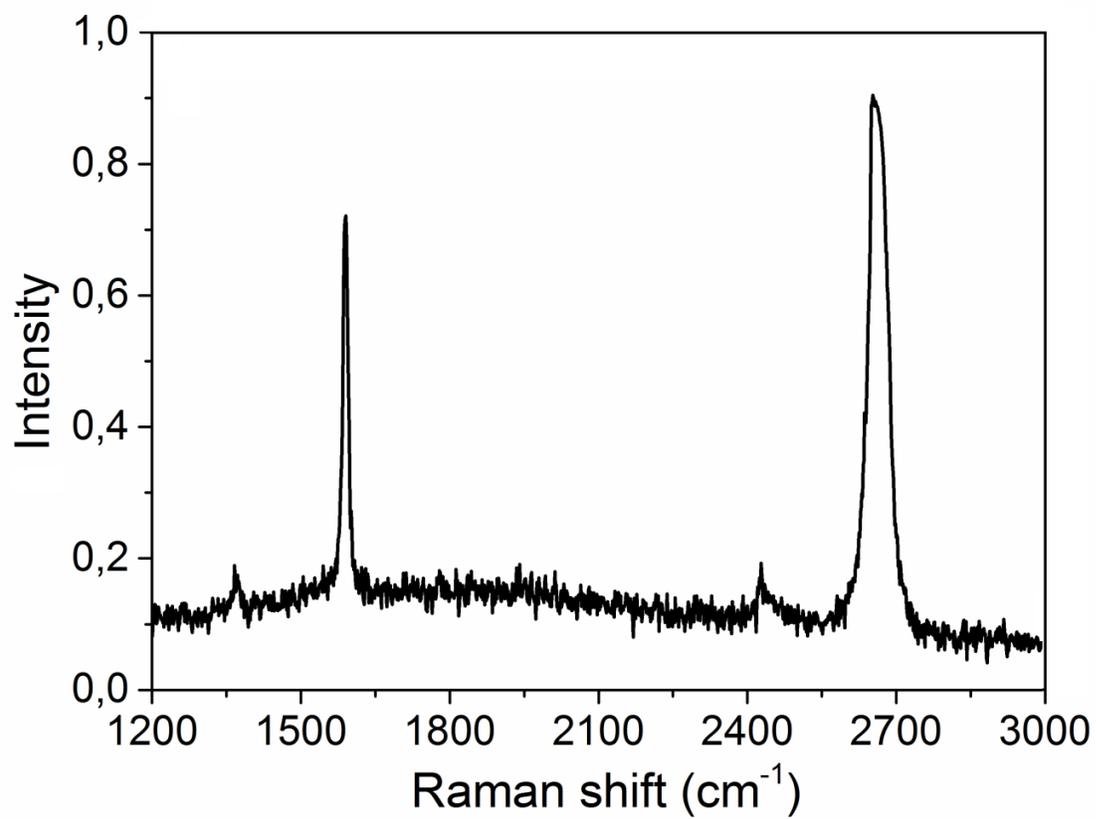

**Fig. S1.**

Raman spectrum of suspended monolayer graphene. The same sample was used in the experiments of Fig. 1.

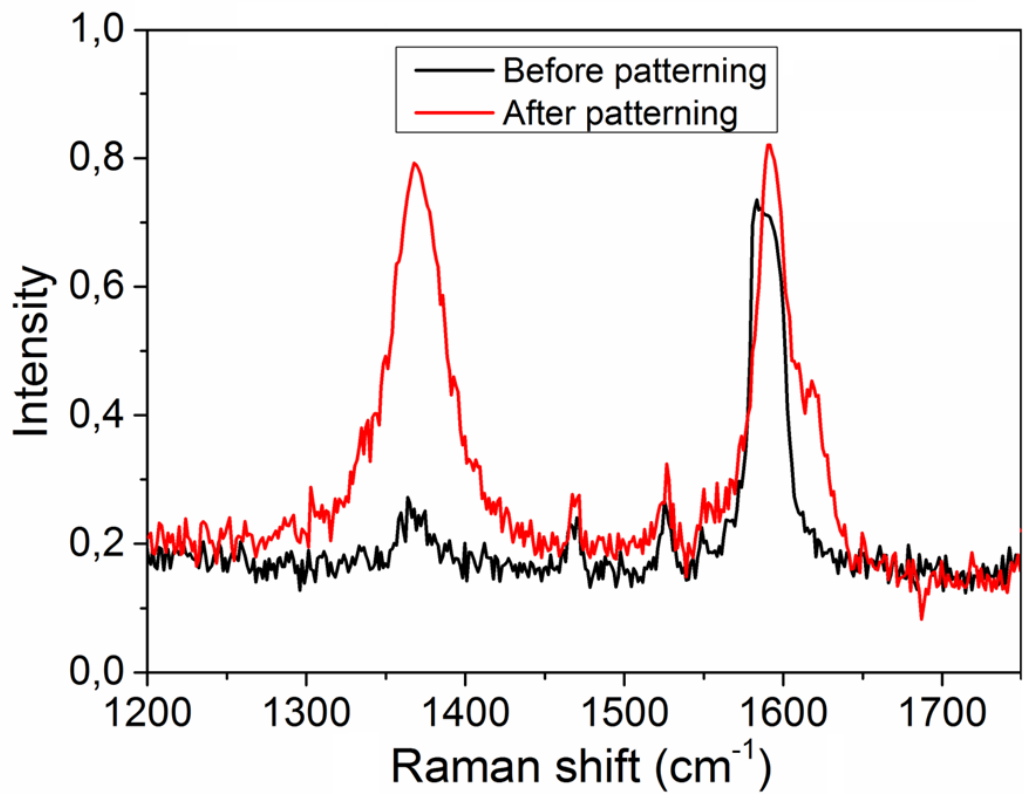

**Fig. S2**

Raman spectrum of monolayer graphene. Black spectrum is from unprocessed graphene and red spectrum shows the Raman spectrum after patterning. Patterning induces very strong growth of the D-band, and the G-band shifts up and changes shape. These changes are key signatures of oxidation of graphene.

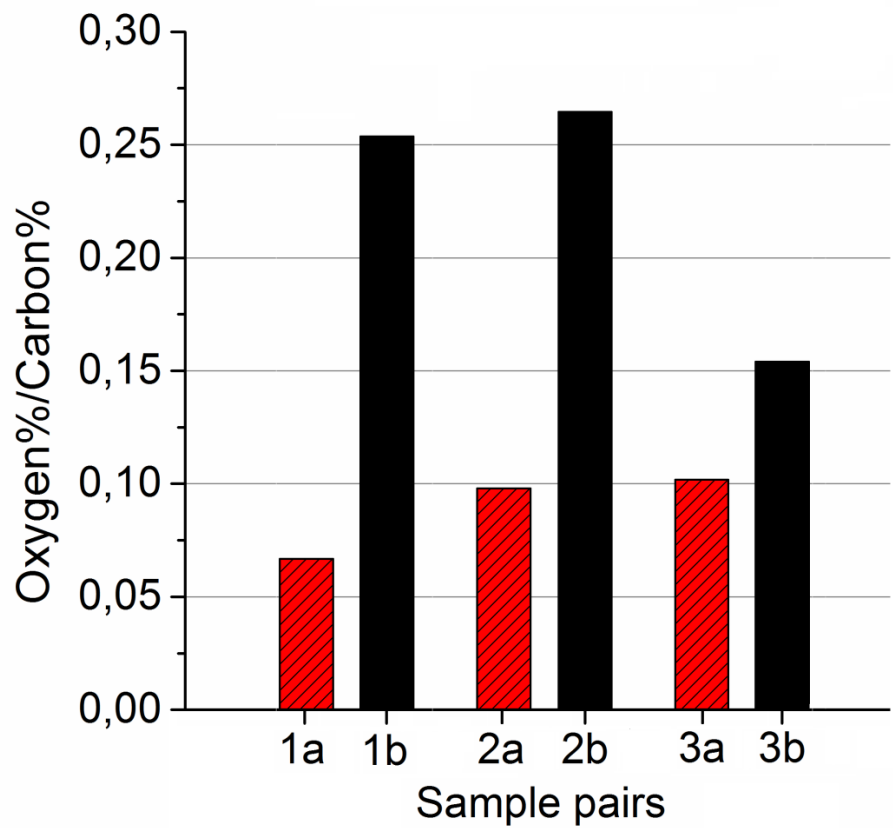

**Fig. S3**

Bar graph of oxygen to carbon ratio in graphene determined by SEM-EDX elemental analysis of air-suspended monolayer graphene. The red bars (a) refer to non-irradiated graphene and black bars (b) correspond to irradiated areas. The measurements for both samples within each sample pair were performed in identical measurement conditions. The oxygen mole fraction increases consistently from ~below10 % up to ~20 %.

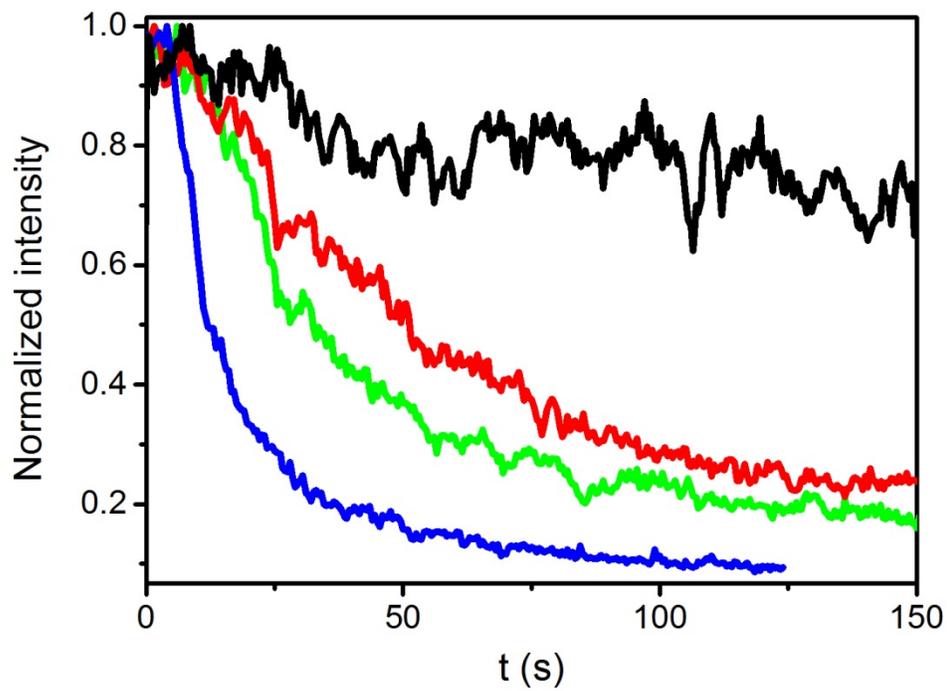

**Fig. S4**

Decay kinetics of the FWM signal upon oxidation. The average power in µW is 8 (black), 10 (red), 13 (green) and 16 (blue).

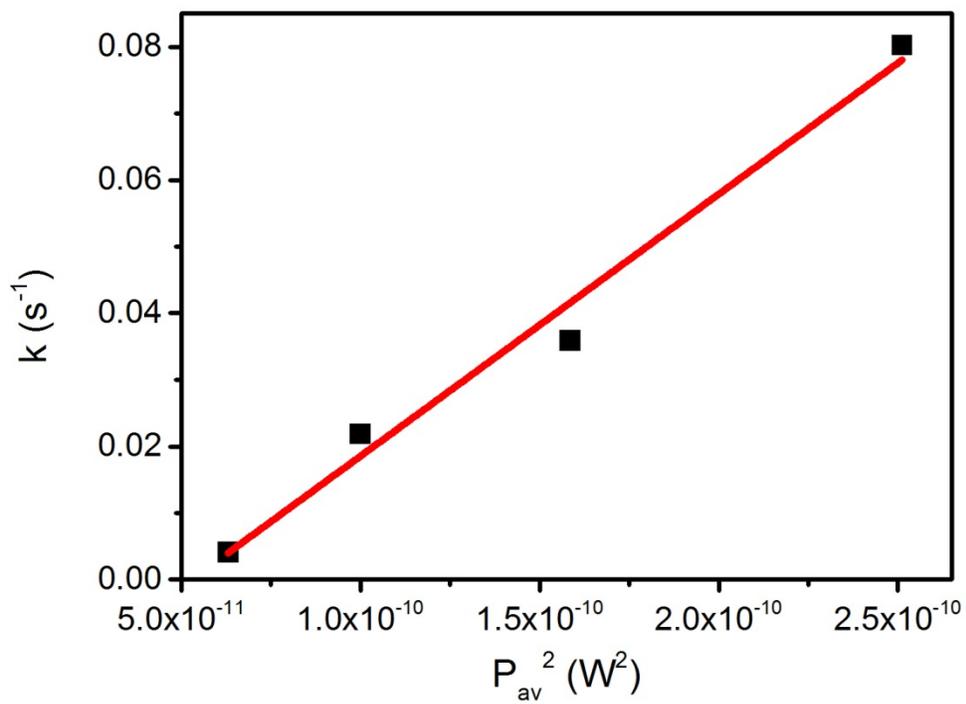

**Fig. S5**

FWM signal decay constant as a function of laser power squared. Linear fit indicates second order behavior, i.e. two-photon process.